\documentclass{vgtc}                          

\graphicspath{{figures/}{pictures/}{images/}{./}} 
\usepackage{amsmath}
\usepackage{amssymb}
\usepackage{times}                     

\usepackage{tabu}                      
\usepackage{booktabs}                  
\usepackage{lipsum}                    
\usepackage{mwe}                       

\usepackage{mathptmx}                  
\UseRawInputEncoding

\onlineid{0}

\vgtccategory{Research}

\vgtcinsertpkg

\title{Co-Designing a Knowledge Graph Navigation Interface: A Participatory Approach}

\author{Stanislava Gardasevic\thanks{e-mail: gardasev@hawaii.edu}\\ %
        \parbox{1.4in}{\scriptsize \centering  Hamilton Library\\ University of Hawaii at Manoa} %
\and Manika Lamba\thanks{e-mail: manika@ou.edu}\\ %
     \parbox{1.4in}{\scriptsize \centering  School of Libraray and Information Studies\\ University of Oklahoma} %
\and Jasmine S. Malone\thanks{e-mail: jsmalone@towson.edu}\\ %
 \parbox{1.4in}{\scriptsize \centering  Albert S. Cook Library\\ Towson University}}  

\teaser{
  \centering
  \includegraphics[width=\linewidth]{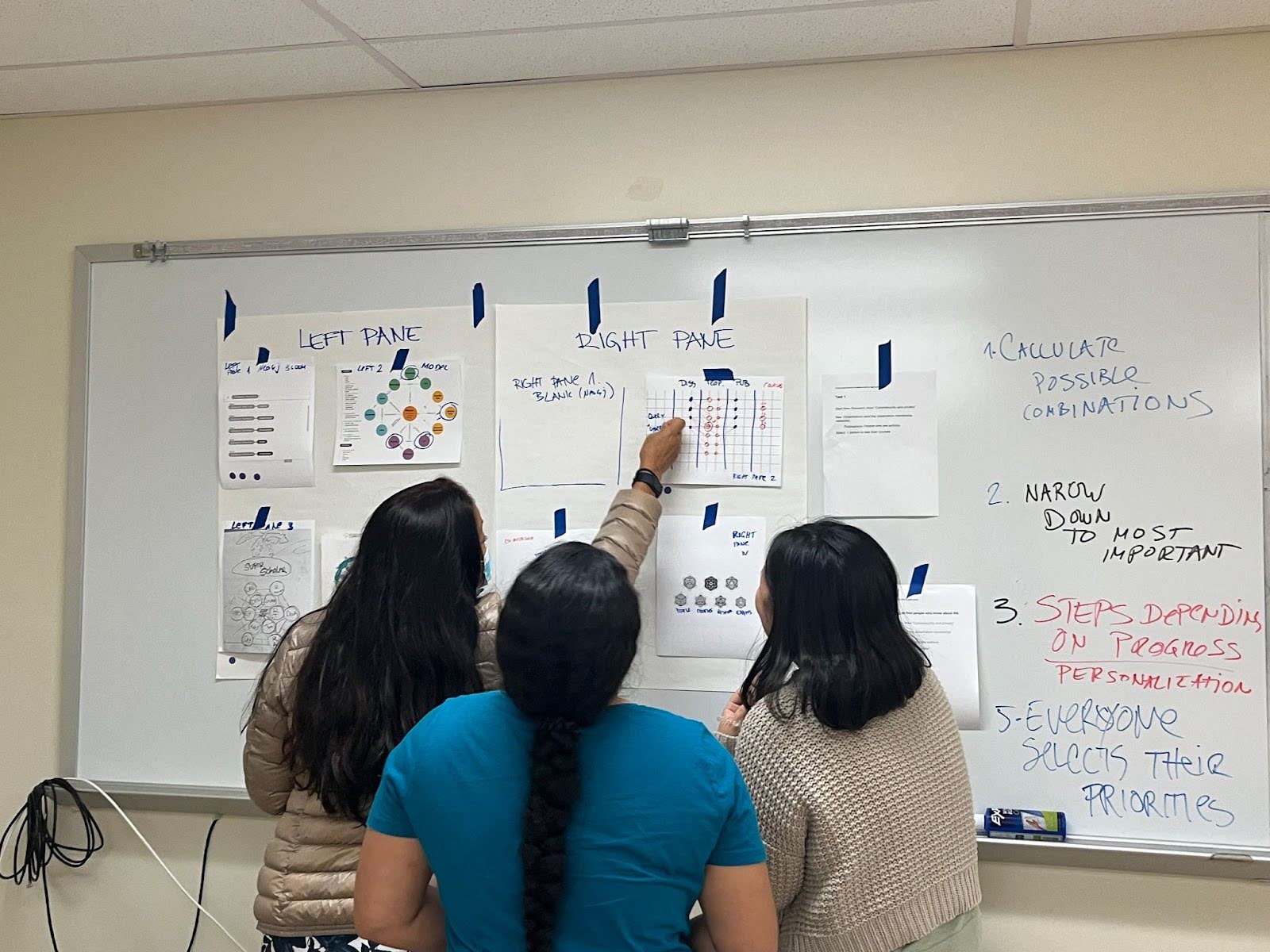}
  \caption{Participatory Design Workshop- Voting for User Interface Left/Right Pane Options}
  \label{fig:1}
}

\abstract{
    Navigating and visualizing multilayered knowledge graphs remains a challenging, unresolved problem in information systems design. Building on our earlier study, which engaged end users in both the design and population of a domain-specific knowledge graph, we now focus on translating their insights into actionable interface guidelines. In this paper, we synthesize recommendations drawn from a participatory workshop with doctoral students. We then demonstrate how these recommendations inform the design of a prototype interface. Finally, we found that a participatory iterative design approach can help designers in decision making, leading to interfaces that are both innovative and user-centric. By combining user-driven requirements with proven visualization techniques, this paper presents a coherent framework for guiding future development of knowledge-graph navigation tools.

} 

\keywords{Participatory design, knowledge graph, PhD students, visualization.}

\begin{document}


\firstsection{Introduction}

\maketitle

Knowledge graphs (KGs) have become a widely utilized technology that can be used to model, capture, and analyze relationships between real-world entities, often representing people and their activities in a social setting. KGs are types of networks in which entities of interest are represented as nodes, and relationships are represented as edges of a network \cite{hogan}. Most academic knowledge graphs—particularly those used in expert recommender systems—rely on a limited set of dimensions, primarily shared research topics, co-authorship, and publication venues, all derived from publication metadata (e.g., \cite{cabanac}). Researchers suggest that a minimum of three dimensions is needed to fully understand the complexity of a social structure represented by the graph \cite{dickison}. Those complex types of KGs are referred to as multilayered or multiplex networks, where the same sets of nodes may differ in each layer \cite{aleta}.

Although visualizing and navigating such complex networks was tackled by many researchers \cite{agarwal, mcgee, nobre, rossi, sun}, this problem is not yet settled.  Visualization of such networks often produces ‘hairball-like’ structures that are difficult to navigate or cause distrust with the end user  \cite{li}. To address this issue, McGee et al. \cite{mcgee} have suggested rephrasing user needs and data as a multilayered network problem while being immersed in the application domain. Our research approach follows this suggestion, addressing the gap in the literature. Therefore, we chose an appropriate case, embedded ourselves in the end-user community, and involved them in each step of the research. Primarily, the participants were actively contributing to the design and population of a multilayered graph dataset in order to fully capture their needs, then we involved them in the ideation for the graph visualization. The authors also did market research identifying the need for such a visualization tool, specifically in academia, after being selected for the NSF I-Corps program in 2023. We spent October of that year conducting market research in two universities with 20 key stakeholders—\textit{graduate students, faculty, program advisors, IT staff, administrators, and postdocs}—each serving as potential end users, decision-makers, or influencers. Their feedback confirmed a strong demand for a centralized knowledge-management system to track both research outputs and teaching activities.

\subsection{Study background}

The case where we applied this research is the interdisciplinary PhD program in \href{https://www.hawaii.edu/cis/}{Communication and Information Sciences (CIS)} at the University of Hawai'i at Mānoa, and the students of this program are the end-users of the KG-based information system. The CIS program has about 30 students, over 40 participating faculty members from four different research area units, and over 100 alumni. The interdisciplinary nature of this program and the diversity in research approach make it a good domain for creating a rich knowledge graph dataset with many facets/dimensions to it, such as departmental affiliation, research area, domain of application, methodological approach, and similar. The interdisciplinary nature of the program allows students to have uniqueness in their research approach; however, there is not enough homogeneity among the student community, which makes it hard for them to get valuable information and insights from their peers \cite{p1}. Still, the students are very much reliant on the word-of-mouth information they get from others \cite{p1,wood} as well as information on courses, exams, and faculty that is currently scattered on different web locations \cite{p1}. For this program, it is expected that the PhD students attend a certain number of courses, pass three qualifying exams, pick a dissertation supervisor, publish a paper while collaborating with supervising faculty members, form an interdisciplinary research committee, defend dissertation proposals, and finally, conduct research and defend the dissertation- all in a certain timeframe (as listed in the program timeline). All of these milestones are taken in account when designing the knowledge-graph based information system, next to other features identified as needed, such as i) support for community building and tacit knowledge exchange in the community, ii) aggregating relevant publicly available information from different web locations (i.e. student and faculty research activities outputs), and iii) gather student generated information on their activities that will help others in decision making, and iv) facilitate discovery of potential interdisciplinary collaborators. The multiplex graph model that resulted from this research has 12 dimensions through which a \textit{Person} can be described or affiliated with other people in the domain (as described in detail in Gardasevic and Gazan \cite{p2}).

Previous stages of the study involved CIS students in interviews and several workshops to inform the design and populate the KG  in order to make a highly usable dataset for the population. This part of the study addresses KG visualization and navigation. Therefore, the guiding question for this paper is: \textbf{what are the end-users’ recommendations for designing the navigation of a knowledge graph information system?}

\section{Literature Review}
Co-designing information and technology solutions with end-users has been well-defined and studied since the mid-twentieth century. Participatory design is a research methodology that seeks to co-design with users by understanding their ways of life \cite{spi} Designing human-centered knowledge resources with doctoral students incorporates this definition across academic disciplines to address underlying information exchange issues within PhD programs \cite{aghaee, cook, deng, p3}. Common problems doctoral students face during their programs are noted as lack of instructor-student interaction \cite{aghaee, p3}, student isolation, time constraints, lack of awareness of rules and regulations, and an imbalance of on-campus and distance learners \cite{aghaee}.  
 
Participatory design research of doctoral students' grounds study in co-designing information systems that influence students’ educational and research journeys. The studies cited sought to ease doctoral student information exchange burdens through technology solutions that motivate student participation both synchronously and asynchronously, for distance learners and on-campus learners \cite{aghaee}. Authors have created Journey Maps \cite{cook}, Knowledge Graphs (KG) \cite{deng,p2}, Information and Communication Technology Support Systems (ICTSS) \cite{aghaee} and Online Doctoral Student Community \cite{naughton} to visualize student identity related to their discipline and program requirements as collaboratively designed knowledge-sharing solutions. The inclusion and guidance of communities in knowledge production better addresses gaps other methods of qualitative study neglect, with researchers evaluating and incorporating their own experiences related to the study \cite{bergold, celi} .

The development of information systems offers students resources and "communication channels" \cite{aghaee} for tacit knowledge exchange: the transference of implicit knowledge often only accessible to a community insider \cite{p3, spi}. In the case of PhD students, knowledge organization systems (KOSs) provide an information discovery opportunity for learners seeking guidance through the relationships presented between peer networking, advisors, and course recommendations to fulfill program requirements \cite{hogan,p4}. Women, first-generation, and underrepresented minority students often have a greater need for insights into the hidden curriculum of graduate education but less access to the social networks where such knowledge is shared \cite{wood}. This benefits research around the provenance and evolution of meaning to information maintained within a KOS \cite{choi}, which further enriches the data presented and heightens its value over time.  

KGs offer visualization of multiplex networks – an overlay of nodes presenting more than one layer of complexity \cite{rossi}. Multi-layered networks present visual challenges to interpreting the data and its overarching relationships, and researchers \cite{mcgee} suggest the need for “interdisciplinary workshops and seminars” to evaluate the usefulness of network complexity to directly address user needs. Others note that knowledge networks can increase social interaction between users within close proximity but potentially decrease the need for other information-seeking activity \cite{phelps}. Li et al. \cite{li} acknowledge the lack of literature describing knowledge graph users' sentiments on information system design, though their research prioritizes researchers, not end-users, thus identifying the research gap of collaborative KG design with the user population, particularly doctoral students. Despite the body of research supporting the need for participatory KG design \cite{aghaee, cook}, there is still misalignment between PhD students' educational goals and information resources that help achieve them, which the co-design of information systems can address.  

\section{Research Methods}

The work described in this paper is part of a larger, longitudinal study that involved end-users in all stages of information system design. Previous papers describe in detail different stages of the study, such as i) requirements gathering \cite{p1}, (ii) KG design \cite{p2}, and iii) graph navigation and evaluation- as presented in Neo4J, a graph database management tool used to populate and explore the CIS KG \cite{p3}. In this paper, we focus on one of the participatory design workshops that were conducted in the later stage of the study, with emphasis on tasks pertinent to designing the novel end-user interface of a proposed information system. 

This workshop was conducted in November 2022, and the first author, an insider in the community, invited the current CIS PhD students via the student email list to participate. The workshop was scheduled during the regular weekly meeting time for the group’s interdisciplinary seminar, and even though the attendance obligation was waived for this session, it was attended by 15 CIS PhD students, with eight participating in person and seven online via Zoom. In this faculty-free zone, two researchers handled different modes of participation and facilitated the discussion.

Prior to splitting into two groups (in-person and online), a researcher provided an introductory overview of the study, discussed issues that arose when visualizing the multilayered graphs, and introduced the tasks participants were expected to perform. Also, there was a joint discussion on the useful graph exploration features noted on the whiteboard (see \autoref{fig:1}, right-hand side). Among other tasks, some of which were different for the two groups (see Gardasevic and Lamba \cite{p3} for more details), the participants had a chance to \textbf{vote }for various options that could be used to navigate the multilayered graphs. They were presented with four different options for the \textbf{left pane}, which was dedicated to selecting categories from the graph, as well as for the \textbf{right pane}, which was used to depict a space where the selected nodes would be projected. Also, participants were presented with a relatable use scenario for graph exploration, which was displayed next to the voting panel. While in-person participants could vote by placing dots on their preferred option as they interacted with one another (see \autoref{fig:1}), online participants had access to a \href{https://app.mural.co/}{Mural} board where they could cast their votes.

Following the workshop, we disseminated information regarding the voting process and other pertinent tasks to the broader CIS student community via the official student email list, thereby enabling those who were unable to attend the workshop to participate and express their views through voting.

Both modes of the workshop were recorded, transcribed using Otter.ai, and analyzed independently by two researchers using \textit{Atlas.ti}. We applied the same code scheme that is relevant to the overall study. The coders reached consensus through discussion on the qualitative coding process. This paper focuses on the findings related to the  "interface design suggestions'' code, outlining novel graph navigation design recommendations that resulted in a prototype.

\section{Results and Design Recommendations}
The results elicited from this study provide valuable insights as we engaged a niche population with high levels of information literacy skills and knowledge in computer and information science pertinent areas.

\begin{figure}[tb]
 \centering 
 \includegraphics[width=\columnwidth]{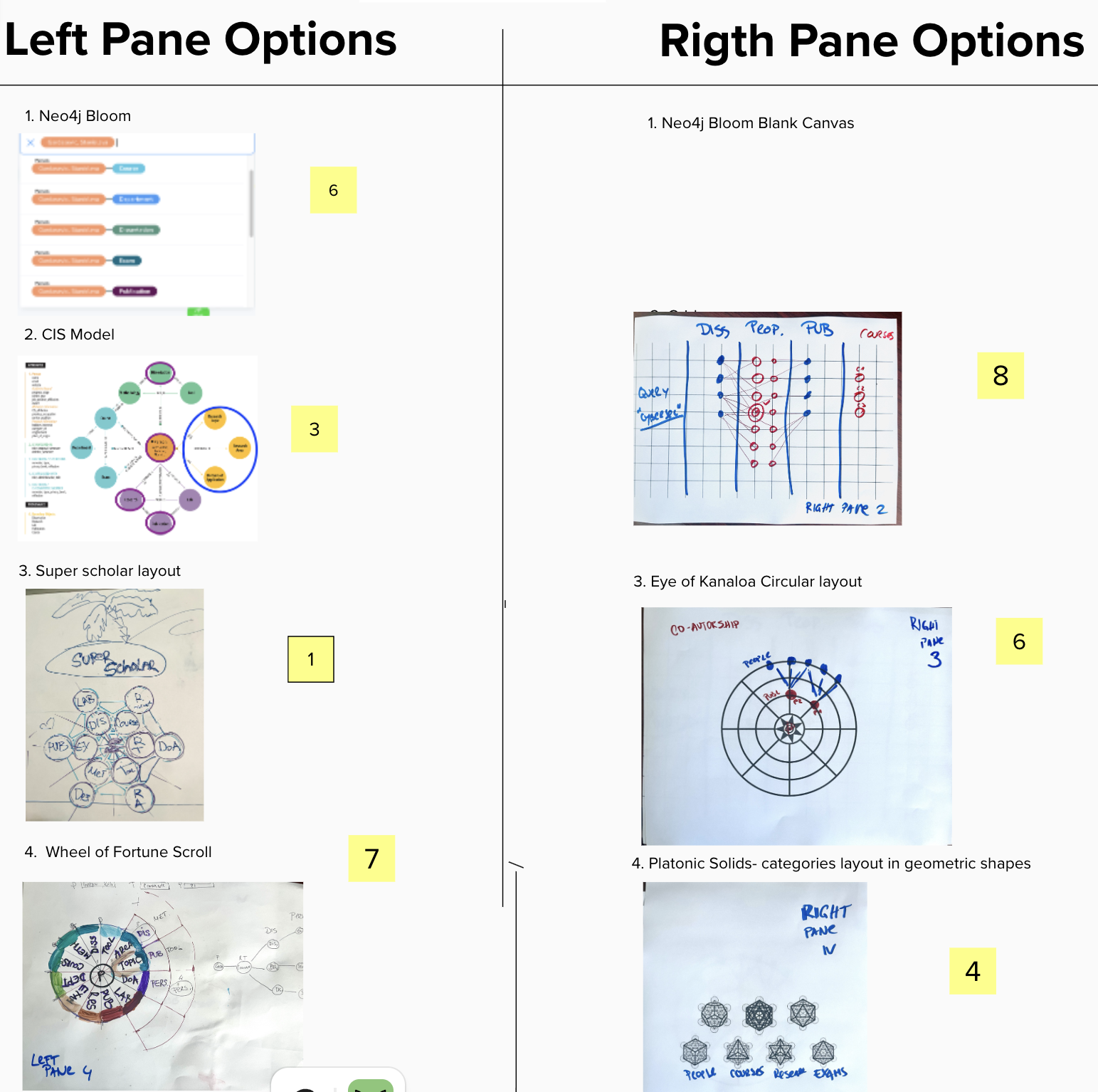}
 \caption{Results of the left/right pane graph navigation voting task}
 \label{fig:2}
\end{figure}

The workshop voting exercise results of both in-person and online participants were consolidated and shown in \autoref{fig:2}. The starting point for understanding the topic of multilayered graph navigation, as demonstrated in the workshop, is depicted as \textit{Option 1} in the \textbf{left pane}, which is the current user interface of Neo4J Bloom, a graph database visualization tool. Even though this option was voted as the second most popular, in the graph evaluation study \cite{p3}, this interface hasn’t performed sufficiently well for the novice end user, requiring a technical background and Cypher querying capabilities.

A valuable suggestion from the group was that the CIS graph-based exploration should heavily rely on recommendations based on students’ self-declared categories of interest and program progress stage, which would limit the search/browse screen and showcase primarily relevant categories, such as people and their activities.

\subsection{Design Recommendations}

The results of this part of the study are combined with insights from the overall study and the findings of the qualitative research by Li et al. \cite{li}, which involved various stakeholders interested in knowledge graph technology, outlining suggestions for visualization. We present a set of design recommendations specifically developed for this multilayered knowledge-graph-based system, which are also broadly applicable to similar systems, particularly those within the educational domain.

\begin{enumerate}
    \item \textbf{User Profile:}  This functionality allows students and faculty to populate and update the categories of their interests, as they may evolve. This feature would allow the recommendation of the most relevant results from the graph, which was suggested as a way to reduce graph visualization complexity. The profile would allow for an automatic or manual progress check feature, so students can track their progress and plan for future steps (contingencies), based on information they see on their peers.

     \item \textbf{Results Recommendations:} Primarily contingent on the entries in the user profile, students should be able to further filter the categories of their interest to get recommendations on relevant students/alumni (to see their pathways, activities such as publishing venues, courses taken), faculty (to explore their supervising engagements, research outputs, collaborations), classes/exams and dissertations/research/publications.

     \item \textbf{Search Functionality:} To support goal-oriented or specific curiosity, a particularly successful approach that emerged from this study was the access to a list of predefined queries that other students had rated as helpful in a different voting exercise, conducted at the same workshop \cite{p3}.

     \item \textbf{Network Visualization Style:} To support open-ended and exploratory search, and in order to significantly reduce the complexity of the graph view often encountered with currently available tools (e.g., Gephi, Neo4J Bloom), we designed a prototype based on the workshop voting results. This prototype design aligns with findings of Li et al. \cite{li}, suggesting that the network could be projected on a layout- in this case, a grid in the right pane- with a reduced number of nodes shown, but with the ability to “see all” or filter based on set criteria. The \textbf{prototype video} (\href{http://bit.ly/3Gxnydo}{http://bit.ly/3Gxnydo}) and \autoref{fig:3} show the novel interface design, created in Figma. Li et al. \cite{li} recommend minimizing the display of complex network structures in knowledge graph visualizations, as the resulting “hairball” effect can confuse and overwhelm users. Instead, they propose using “knowledge cards” to present summarized information about nodes. Building on this approach, we suggest generating a knowledge card for each faculty member that highlights key activities and relevant statistics, such as the number of dissertations chaired or participated in and the average duration of their students’ studies. These cards would also provide options to display collaborators from both the CIS faculty and student communities.
     \item \textbf{Nodes and Edges Layout:} Color coding is important to make a clear distinction between people’s affiliations with different departments or disciplines, their role (faculty vs. students), and student status (current vs. past students). The same goes for the different types of relationships/edges that are linking the same nodes. For instance, \[(Person)-[\text{is\_teaching vs. has\_taken}] \rightarrow (\text{Course})\]. However, this is an interdisciplinary domain, meaning there are many colors in the visualization, which requires a color legend overview available at all times. Finally, the underlying schema should be available for a quick overview to support the exploration of the graph based on many possible paths for each node \cite{li}.
    \item \textbf{Showing temporal dimension:} The temporal dimension of the graph is very important for this and other user populations \cite{li}, yet Neo4J Bloom does not allow sufficient support
    for a layout based on the \textbf{timestamps}. Except for the planning aspect (where one should be able to see the animation of personal graph evolution), it is important to show the timestamp for the
    course teaching/taking activities, as this relationship allows for attributes to capture qualitative information on the student’s experiences at a course, and those may evolve over time.
    
\begin{figure}[tb]
 \centering 
 \includegraphics[width=\columnwidth]{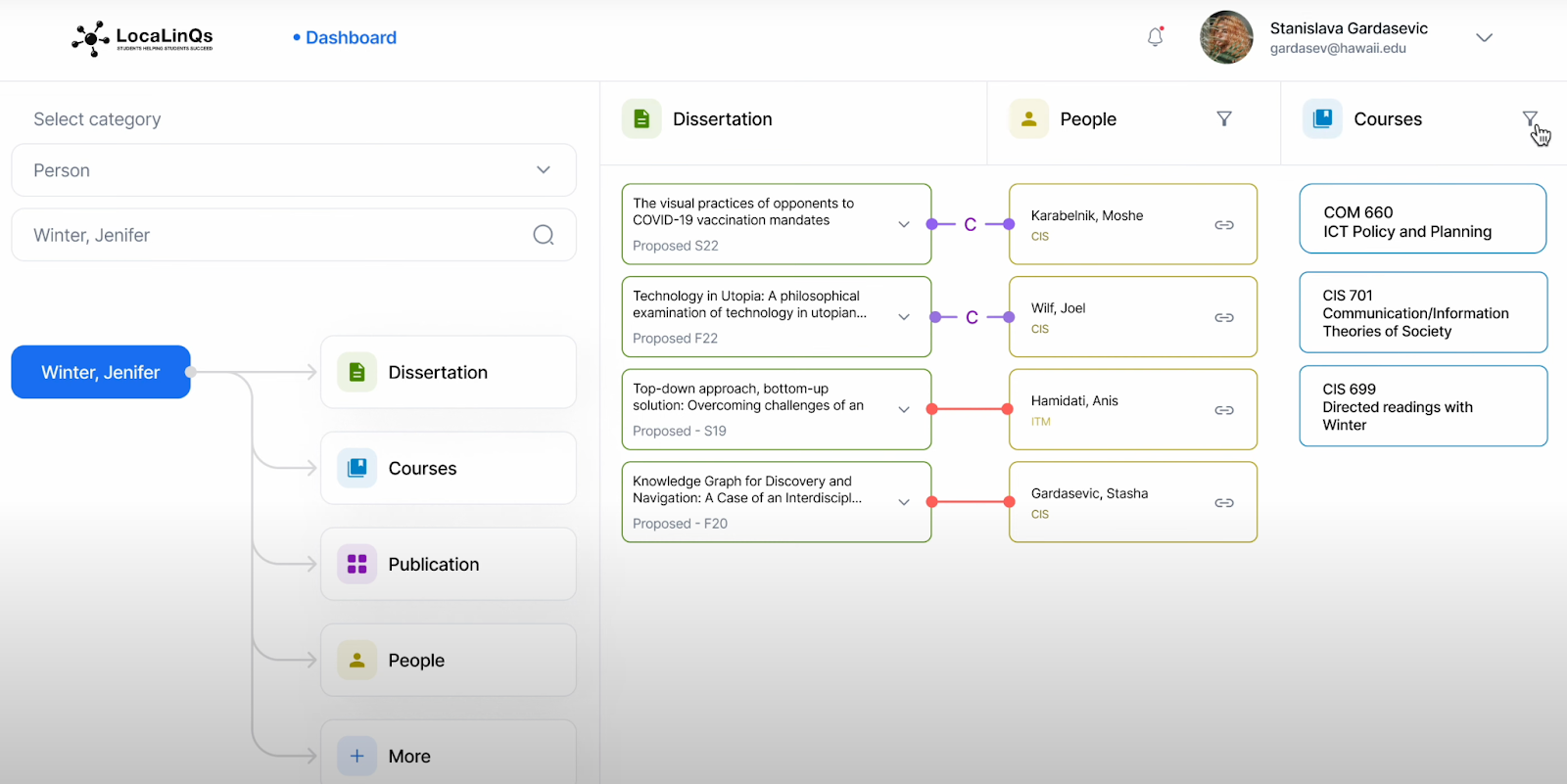}
 \caption{Graph Navigation Interface Prototype Screenshot}
 \label{fig:3}
\end{figure}

\end{enumerate}

\subsection{Limitations and Future Research}

Although this paper proposes a human-centered design for a multi-layered graph visual exploration tool, the prototype has not yet been developed into a fully functional one, which would allow us to evaluate it with end users. We plan to develop it further to include all of the above design recommendations.

Therefore, we aim to extend this research by integrating the Graph Retrieval-Augmented Generation (GraphRAG) workflow into the knowledge graph information system population and interface, combining the structural precision of knowledge graphs with the semantic flexibility of large language models (LLMs). Preliminary findings indicate that GraphRAG effectively builds or updates graphs directly from raw textual documents, facilitates targeted retrieval along meaningful entity–relation paths, and leverages graph-anchored snippets to significantly reduce hallucinations in LLM-generated summaries and answers \cite{potts, amazon}. Such integration would enable users to naturally articulate queries. For example, ``Who in our program has published on \textit{explainable AI}?'' and receive concise, provenance-linked responses while retaining the ability to visually explore the underlying graph structure.

Multiple cloud-based and open-source solutions now considerably lower the technical barrier for adopting GraphRAG prototypes into production environments. These include Amazon Bedrock Knowledge Bases featuring managed GraphRAG with Neptune Analytics \cite{amazon}, the AWS open-source GraphRAG Toolkit \cite{aws}, Microsoft Research’s modular GraphRAG pipeline \cite{micro}, and Neo4j’s dedicated GraphRAG Python package \cite{neo}. Moving forward, we plan to perform participatory design with PhD students and faculty to get insights into the updated prototype of the knowledge system embedded with GraphRAG to observe how users trust, refine, and contribute to an AI-augmented graph. This phase will help us to find new heuristics for transparency, feedback, and continual graph curation in scholarly knowledge-work interfaces.

\section{Conclusion}
A key lesson from this study is the value of participatory design, especially in an academic setting. We engaged PhD students throughout the design process via a series of hybrid workshops, where initial prototypes and data models were iteratively refined based on their feedback. This inclusive approach proved critical for uncovering tacit knowledge needs and ensuring the interface resonated with actual user workflows. Creating a faculty-free, open forum for students (with measures for anonymity) encouraged honest input regarding pain points and desired features. The hybrid format expanded accessibility, though it required careful facilitation to maintain coherence across sessions. Methodologically, our approach responds to the gap noted by Li et al. \cite{li}, who observed that many knowledge graph tools are designed in isolation and “disconnected from the communities they are supposed to serve”. By contrast, involving users early and often grounded our design decisions in the lived experiences of the community. We learned that participatory co-design not only produces a more usable interface but also builds user trust in the system’s relevance and accuracy, an important factor for the adoption of knowledge graph tools. These insights reinforce that future visualization research should treat end-users as partners in the design process. Particularly in academic or specialized domains, a participatory, iterative design approach can surface requirements that designers alone might miss, leading to interfaces that are both innovative and user-centric.

\acknowledgments{
The authors would like to acknowledge the invaluable contributions of all the CIS PhD students and alumni who participated in this research.}



\end{document}